\documentclass[pra,twocolumn,superscriptaddress,10pt,floatfix]{revtex4}
\usepackage[pdftex,plainpages=false,colorlinks=true,citecolor=blue,linkcolor=blue,urlcolor=blue,filecolor=green,bookmarksopen=true]{hyperref}
\usepackage[utf8]{inputenc}
\usepackage[english]{babel}
\usepackage{amsmath}
\usepackage[caption = false]{subfig}
\usepackage{graphicx,epstopdf}
\usepackage{blindtext}
\usepackage{lipsum}
\usepackage{amsfonts}
\usepackage{bbm}
\usepackage{amssymb}
\usepackage{enumerate}
\usepackage{color}
\usepackage{latexsym}
\usepackage{times,txfonts}

\newcommand{\Dcal}{\mathcal{D}}

\newcommand{\Rcal}{\mathcal{R}}

\newcommand{\Ical}{\mathcal{I}}

\newcommand{\1}{\mathbbm{1}}

\newcommand{\Rmath}{\mathbbm{R}}

\newcommand{\ket}[1]{\left| #1 \right\rangle}
\newcommand{\bra}[1]{\left\langle #1 \right|}
\newcommand{\ve}[1]{\langle #1 \rangle}

\newcommand{\tr}[1]{ \text{Tr}\left\{ #1 \right\}}

\begin{document}

\title{Optical simulation of a quantum thermal machine}

\author{M. H. M. Passos}
\email{mhmpassos@id.uff.br}
\affiliation{Instituto de Ciências Exatas, Universidade Federal Fluminense, Volta Redonda, Rio de Janeiro, Brazil}
\affiliation{Instituto de F\'{i}sica, Universidade Federal Fluminense, Avenida General Milton Tavares de Souza s/n, Gragoat\'{a}, 24210-346 Niter\'{o}i, Rio de Janeiro, Brazil}

\author{Alan C. Santos}
\email{ac\_santos@id.uff.br}
\affiliation{Instituto de F\'{i}sica, Universidade Federal Fluminense, Avenida General Milton Tavares de Souza s/n, Gragoat\'{a}, 24210-346 Niter\'{o}i, Rio de Janeiro, Brazil}

\author{Marcelo S. Sarandy}
\email{msarandy@id.uff.br}
\affiliation{Instituto de F\'{i}sica, Universidade Federal Fluminense, Avenida General Milton Tavares de Souza s/n, Gragoat\'{a}, 24210-346 Niter\'{o}i, Rio de Janeiro, Brazil}

\author{J. A. O. Huguenin}
\email{jose\_huguenin@id.uff.br}
\affiliation{Instituto de Ciências Exatas, Universidade Federal Fluminense, Volta Redonda, Rio de Janeiro, Brazil}
\affiliation{Instituto de F\'{i}sica, Universidade Federal Fluminense, Avenida General Milton Tavares de Souza s/n, Gragoat\'{a}, 24210-346 Niter\'{o}i, Rio de Janeiro, Brazil}

\begin{abstract}
We introduce both a theoretical and an experimental scheme for simulating a quantum thermal engine through an all-optical approach, 
with the behavior of the working substance and the thermal reservoirs implemented via internal degrees of freedom of a single-photon.
By using the polarization and propagation path, we encode two quantum bits and then implement the thermodynamical steps of an 
Otto cycle. To illustrate the feasibility of our proposal, we experimentally realize such simulation through
an intense laser beam, evaluating heat and work at each individual step of the  thermodynamical cycle. 
In addition, from the analysis of the entropy production during the entire cycle, we can study the amount of quantum friction 
produced in the Otto cycle as a function of the difference of temperature between hot and cold reservoirs. 
Our investigation constitutes, therefore, an all-optical-based thermal machine simulation and opens perspectives for 
other optical simulations in quantum thermodynamics.
\end{abstract}

\maketitle

\section{Introduction}

The idealization of microscopic quantum systems allowing for extraction of work and heat is at the heart of quantum thermal engines (QTEs), quantum refrigerators~\cite{Geva:92,Lutz:16,Noah:10}, and quantum batteries~\cite{Alicki:13,Binder:15}. In analogy with its classical counterpart, a QTE has a quantum system as its \textit{working substance}, which interacts with thermal reservoirs at different temperatures $\beta_{\text{c}}$ and $\beta_{\text{h}}$. Indeed, a number of works have explored the quantum nature of the working substance in order to investigate whether we can get optimal performance of  QTEs in comparison with their classical counterparts. Concrete proposals of QTEs have been studied in recent years from different approaches~\cite{Kieu:04,Scully:03,Kosloff:84,Rezek:06,Henrich:07,Geusic:59,Quan:05,Lutz:16-Science,Scovil:59}.  Moreover, experimental verifications of the performance of a QTE 
have recently been achieved, e.g., nuclear spin systems manipulated through nuclear magnetic resonance (NMR)~\cite{Peterson:18} and nitrogen-vacancy centers in 
diamond~\cite{Klatzow:17}. In general, a major difficulty for implementing QTEs in real physical systems is the high controllability required so that robustness against decoherence is achieved. Therefore, there is great interest in designing QTEs from architectures that offer efficient control of reservoirs.

In order to simulate controllable reservoirs, we have to consider the effect of quantum channels acting on quantum information~\cite{Nielsen:Book}. In this context, it is 
fundamental in our approach to take into account the optical implementation of relevant quantum channels, such as amplitude damping, phase-damping (PD), and bit flip channels, among others performed by using single photons~\cite{PRA.78.Davidovich}. On the other hand, degrees of freedom of an intense laser beam have been widely used to simulate single-photon experiments and the results show that such procedure consists in a relevant test-bed for several quantum properties in a rather simple way~\cite{PRL.99.Topo}. 
Indeed, it can be shown that such systems can be used to observe violations of Bell's like inequality~\cite{PRA.82.Borges,NaturePhot.Kagalwala-2012} and Mernin's inequality for tripartite systems \cite{Opt.Lett.Balthazar-2016}.  Moreover, many other quantum protocols can be investigated, such as quantum key distribution~\cite{PRA.77.Cadu-Cripto}, teleportation~\cite{PRA.83.Zela-Teleport} and quantum logical gates~\cite{Opt.Exp.Souza-Cnot-2010,JOPS-B.Cod.Op.Balthazar-2016}. As a further implementation of interest here, it is important to highlight the experimental simulation of open quantum systems to investigate environment-induced entanglement~\cite{PRA.97.Enviro-Passos}.
In this paper, we propose an all-optical-based scheme, which allows us to simulate the performance of a thermal machine in quantum mechanics and perform an experimental simulation by using degrees of freedom of an intense laser beam. 
We theoretically show how we can construct a quantum machine by using the dephasing channel to mimic a thermal reservoir. 
We then  implement the Otto cycle for polarization of a single-photon through a simulation via linear optical circuit.

 \section{ Quantum thermal machine}

\subsection{Quantum Otto cycle}

Let us begin by the definition of heat and work in quantum mechanics. In general, heat and work are not 
quantum observables~\cite{Talkner:07}. However, 
for the processes of interest here, either heat or work will be vanishing. In this situation, convenient expressions can be derived from the first law of thermodynamics. 
Indeed, by considering the internal energy  $U(t)$ of a quantum system described by a density operator $\rho(t)$ at instant $t$ as $U(t) = \tr{H(t)\rho(t)}$, with 
$H(t)$ denoting the Hamiltonian of the system, it is possible to define \textit{work} $\delta W(t)$ and \textit{heat} $\delta Q(t) $ for infinitesimal processes as~\cite{Alicki:79,Anders:13}
\begin{eqnarray}
\delta W(t) & = & \tr{\dot{H}(t)\rho(t)}dt \text{ , } \label{dW} \\
\delta Q(t) & = & \tr{H(t)\dot{\rho}(t)}dt \text{ . } \label{dQ}
\end{eqnarray}
Notice that $\delta W(t) > 0$ ($\delta W(t) < 0$) implies that work is being performed on (by) the system, so that its internal energy is increasing (decreasing). 
Similarly, when $\delta Q(t) > 0$ ($\delta Q(t) < 0$) we say that heat is being injected in (extracted from) the system. More details are presented in Appendix~\ref{ApHeatWork}.

\begin{figure}[t!]
	\centering
	\includegraphics[scale=0.7]{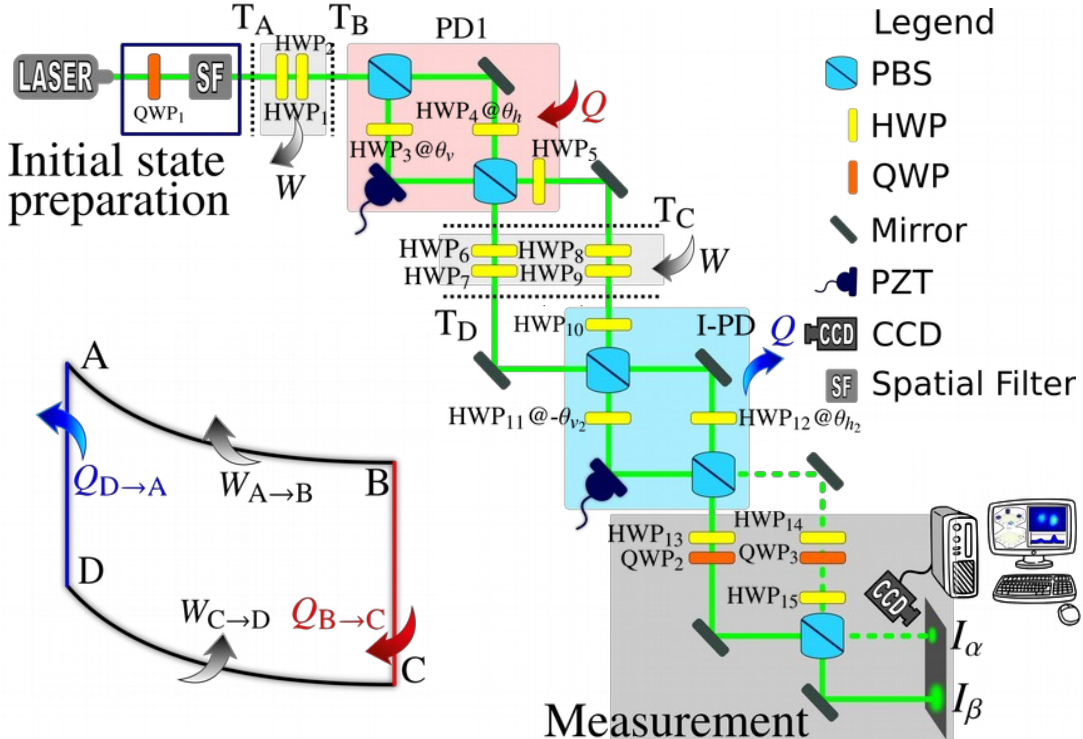}
	\caption{Experimental circuit for implementing the Otto cycle, with each step of the thermodynamical cycle identified in the experimental setup.}
	\label{Fig1-2}
\end{figure}

We will consider a quantum thermal machine realizing an Otto cycle, whose steps are shown in Fig.~\ref{Fig1-2} and can be described as follows.
\textit{Gap expansion step --} Initially a quantum bit (qubit) is prepared in a thermal state of the reference Hamiltonian $H_{\text{e}}(0) = \omega_{\text{ini}} \sigma_{y}$, at inverse temperature $\beta_{\text{c}}$. Thus, the system undergoes a unitary dynamics driven by the Hamiltonian $H_{\text{e}}(t) = \hbar \left[ \omega_{\text{ini}} f(t) + \omega_{\text{fin}} g(t) \right]\sigma_{y}$, with functions $\{g,f\}:t\in\Rmath \rightarrow g,f\in\Rmath$ satisfying $g(0)=f(\tau) = 1$ and $g(\tau)=f(0) = 0$ and $|\omega_{\text{ini}}|<|\omega_{\text{fin}}|$.  This constitutes the branch $A\rightarrow B$ of the thermal engine cycle  in Fig.~\ref{Fig1-2}, where an amount of work $W_{\text{A}\rightarrow\text{B}}$ is performed by the engine.
\textit{Thermalization with hot reservoir --} At this stage, the system is coupled to a thermal reservoir at inverse temperature $\beta_{\text{h}}$, thermalizing with it. Therefore, the final state at this step is a thermal state of the Hamiltonian $H_{\text{e}}(\tau)$ at inverse temperature $\beta_{\text{h}}$. In this step, the system exchange heat with the reservoir, but no work is performed. In this branch $B\rightarrow C$  (see Fig.~\ref{Fig1-2}), heat $Q_{\text{B}\rightarrow\text{C}}$ is introduced into engine.
\textit{Gap compression step --} Now, we switch off the interaction between the system and the reservoir. Thus, we drive the system by a time-dependent Hamiltonian $H_{\text{c}}(t) = \hbar \left[ \omega_{\text{fin}} f(t) + \omega_{\text{ini}} g(t) \right]\sigma_{y}$. In this branch $C\rightarrow D$  (see Fig.~\ref{Fig1-2}), an amount of work  $W_{\text{C}\rightarrow\text{D}}$ is being performed on the system.
\textit{Thermalization with cold reservoir --} To end, the system is coupled to the cold reservoir, in which the final state is the thermal state of the Hamiltonian $H_{\text{c}}(\tau)$ at inverse temperature $\beta_{\text{c}}$. No work is performed, but heat is exchanged between the system and the cold reservoir. This is the last branch $D\rightarrow A$ 
of the cycle  (see Fig.~\ref{Fig1-2}), with an amount of heat $Q_{\text{D}\rightarrow\text{A}}$ extracted from the engine.
From Eqs.~(\ref{dW}) and~(\ref{dQ}) we can compute heat and work for each step of this Otto cycle (shown in Fig.~\ref{Fig1-2}) as
\begin{eqnarray}
W_{\text{A}\rightarrow\text{B}} &=& - \hbar (\omega_{\text{fin}} -\omega_{\text{ini}})\tanh (\hbar \omega_{\text{ini}}  \beta_{\text{c}}) \text{ , } \label{Wab} \\
Q_{\text{B}\rightarrow\text{C}} &=& \hbar \omega_{\text{fin}} \left[ \tanh (\hbar \omega_{\text{ini}} \beta_{\text{c}}) - \tanh (\hbar \omega_{\text{fin}} \beta_{\text{h}}) \right] \text{ , } \label{Qbc} \\
W_{\text{C}\rightarrow\text{D}} &=& \hbar (\omega_{\text{fin}} -\omega_{\text{ini}})\tanh (\hbar \omega_{\text{fin}}  \beta_{\text{h}}) \text{ , }  \label{Wcd} \\
Q_{\text{D}\rightarrow\text{A}} &=& - \hbar \omega_{\text{ini}} \left[ \tanh (\hbar \omega_{\text{ini}} \beta_{\text{c}}) - \tanh (\hbar \omega_{\text{fin}} \beta_{\text{h}}) \right] \text{ , } \label{Qda}
\end{eqnarray}
where we can derive the condition between $\beta_{\text{c}}$ and $\beta_{\text{h}}$ in order to get $Q_{\text{D}\rightarrow\text{A}} < 0 $ as $\omega_{\text{ini}} \beta_{\text{c}} > \omega_{\text{fin}} \beta_{\text{h}}$. Such condition establishes the relation between the parameters of the reservoir and Hamiltonian as $T_{\text{h}}/T_{\text{c}} > \omega_{\text{fin}}/\omega_{\text{ini}}$.

\subsection{Simulation of a thermal reservoir with the phase damping channel}

We can simulate the required reservoirs for implementing a quantum  thermal machine by using a PD quantum channel for a single-photon. To see this, we first need to realize that our system is initially prepared in a thermal state of $H_{\text{ini}}$ at inverse temperature $\beta_{\text{c}}$, which reads 
\begin{equation}
    \label{istate}
\rho_{\text{ini}}^{\text{th}} = \frac{1}{2}\left[ \1- \tanh (\hbar \omega_{\text{ini}} \beta_{\text{c}})\sigma_{y}\right] \text{ . } 
\end{equation}

Thus, due to the contact of our system with a thermal reservoir at inverse temperature $\beta_{\text{h}}$, under action of the Hamiltonian $H_{\text{fin}}$, the state after thermalization will be 
\begin{equation}
\rho_{\text{fin}}^{\text{th}} = \frac{1}{2}\left[ \1 - \tanh (\hbar \omega_{\text{fin}} \beta_{\text{h}})\sigma_{y}\right] \text{ . }
\label{rhofin-th}
\end{equation}
Thus, the reservoir just changes the off-diagonal elements of the initial state, from $\tanh (\hbar \omega_{\text{ini}} \beta_{\text{c}})$ to $\tanh (\hbar \omega_{\text{fin}} \beta_{\text{h}})$. On the other hand, given a density matrix $\rho$ with elements $\rho_{nm}$, we know that a  PD channel acts over the elements $\rho_{01}$ and $\rho_{10}$ as $\rho_{01} \rightarrow \rho_{01} e^{-\gamma \tau_{\text{d}}}$ and $\rho_{10} \rightarrow \rho_{10} e^{-\gamma \tau_{\text{d}}}$, respectively, where $\gamma$ is the dephasing rate and $\tau_{\text{d}}$ is the total time interval of interaction of our system with the decohering reservoir~\cite{Nielsen:Book}. To conclude, by applying this map to the state $\rho_{\text{ini}}^{\text{th}}$ we have $\tanh (\hbar \omega_{\text{ini}} \beta_{\text{c}}) \rightarrow e^{-\gamma \tau_{\text{d}}} \tanh (\hbar \omega_{\text{ini}} \beta_{\text{c}})$, so that we can adjust the parameter $\gamma \tau_{\text{d}}$ to get the parameter $\beta_{\text{h}}$ from
\begin{eqnarray}
\hbar \omega_{\text{fin}} \beta_{\text{h}} = \text{arctanh}\left[e^{-\gamma \tau_{\text{d}}} \tanh (\hbar \omega_{\text{ini}} \beta_{\text{c}}) \right] \text{ . } \label{omegabeta}
\end{eqnarray}
Thus, one can use the phase-damping channel to simulate the thermal reservoir in a heat engine, where we set the parameter $\gamma \tau_{\text{d}}$ 
to encode the hot reservoir temperature.  It is important to mention here that the state $\rho_{\text{fin}}^{\text{th}}$  in Eq.~(\ref{rhofin-th}), obtained  after imposing $\hbar \omega_{\text{fin}} \beta_{\text{h}}$ as given by Eq.~(\ref{omegabeta}), 
does not represent  an actual thermal (Gibbs) state, since we do not have  actual thermal baths 
 in contact with the quantum system. Its correct meaning should be understood as a simulated thermal state, 
 which is achieved by mapping the desired temperature in terms of the dephasing parameters.

In several schemes of QTEs~\cite{Kosloff:84,Rezek:06,Henrich:07,Geusic:59,Quan:05,Lutz:16-Science,Scovil:59}, both steps of compression and expansion are performed by slow (adiabatic) unitary evolution, so that an amount of work is performed on/by the system and no heat is exchanged. However, since any unitary dynamics suppresses the heat exchange  (closed system), we can implement a fast evolution in this step~\cite{Lutz:18}. In single-photon experiments we can simulate the dynamics of a quantum system through unitary operators, thus the expansion and compression steps are implemented by unitary $U_{\text{e}}(\tau)$ and $U_{\text{c}}(\tau)$, respectively, where $\tau$ is the total compression/expansion time interval (adopted to be the same in both steps). By writing the expansion/compression Hamiltonian as $H_{\text{c/e}}(t) = \hbar \omega_{\text{c/e}}(t) \sigma_{y}$, the unitary evolution operator is given as $U_{\text{c/e}}(\bar{\omega}\tau) = e^{-i\bar{\omega}_{\text{c/e}}\tau \sigma_{y}}$, where we denote $\bar{\omega}_{\text{c/e}} = (1/\tau)\int_{0}^{\tau}\omega_{\text{c/e}}(t)dt$. 

\section{Experimental implementation}

In this section we discuss the optical encoding of the Otto cycle discussed above so that we can simulate it with our particular system, with a general schematic representation shown in Fig.~\ref{Fig1-2}. The working substance  and an auxiliary (ancilla) system are encoded in the degrees of freedom of a laser beam. The qubit associated with the machine, in which we will extract/introduce heat and work, are the two independent photon polarization states $\ket{V}$ (vertical) and $\ket{H}$ (horizontal).

\subsection{ PD channel with linear optical circuits}

The experimental implementation of PD channels that simulate the thermal reservoirs has been performed by using linear optical circuits. In our experiment, instead of a single-photon source, we used an intense laser beam that can be described by a coherent state with a macroscopic photon number. This approach has been successfully explored in literature in different scenarios~\cite{PRA.77.Cadu-Cripto,PRA.83.Zela-Teleport,Opt.Exp.Souza-Cnot-2010,JOPS-B.Cod.Op.Balthazar-2016,PRA.97.Enviro-Passos}. For this reason, we will present the experiment by using Dirac notation for polarization states once the discussion for single-photon states is straightforward. We encoded the qubit in the polarization degree of freedom and the environment in the propagation direction (path). For the polarization states, we have a two-level system, where we can associate the horizontal polarization as a ground state ($\ket{H}_{\text{S}}\equiv\ket{0}_{\text{S}}$) and the vertical polarization as an excited state ($\ket{V}_{\text{S}}\equiv\ket{1}_{\text{S}}$). In case of the propagation direction, we encoded the path also as a two-level system, with orthogonal directions, $\vec{k}_0$ and $\vec{k}_1$, representing the reservoir ground ($\ket{0}_{\text{R}}$) and excited state ($\ket{1}_{\text{R}}$), respectively.

The scheme for the PD channel is shown in Fig.~\ref{Fig1-2} (PD1, red square in the circuit). To describe the channel action on the polarization states, let us consider, without loss of generality, an incoming laser beam described by a right-circular polarized state 
\begin{equation}
\ket{\psi_{\textrm{RC}}} = \frac{1}{\sqrt{2}} \left( \ket{H} - i \ket{V} \right) \text{ , }
\end{equation}
which will interact with the reservoir. It is worth mentioning that the density matrix associated with above state is written as $\rho_{\textrm{RC}} = \ket{\psi_{\textrm{RC}}}\bra{\psi_{\textrm{RC}}}=(1/2)(\1 - \sigma_{y})$. Thus, by considering  thermal states such that $k_{\text{B}}T_{\text{c}} \approx \hbar \omega_{\text{ini}}/3$, we can use the approximation $\tanh({\hbar\omega_{\text{ini}}\beta_\text{c}}) \approx 1$ in Eq.~\eqref{istate} to see that we (approximately) get the 
same the density matrix as that for the state $\ket{\psi_{\textrm{RC}}}$. Therefore, as the state $\ket{\psi_{\text{RC}}}$ arrives at the channel, the polarization beam splitter PBS1 transmits (reflects) the horizontal (vertical) polarization state. In this way, the H-polarization component (H) goes to the half-wave plate HWP4@$\theta_h$ ($\theta_h = 0^{\circ}$), where no change occurs in the polarization component (H) of transmitted arm. On the other hand, for the reflected arm, the V-polarization component passes through HWP3@{$\theta_v$}, implementing the transformation
\begin{eqnarray}
\ket{V}_{\text{S}} \rightarrow \sin(2\theta_v)\ket{H}_{\text{S}} + \cos(2\theta_v)\ket{V}_{\text{S}} \text{ . } \label{hwptransform}
\end{eqnarray}
Moreover, in this reflected arm, we introduced a piezoelectric ceramic (PZT) placed in the mirror for adjusting the difference of phase ($\Delta\phi$) between the two arms.
In this way, by adjusting $\Delta\phi = 0$, we have the state of polarization of the transmitted arm going out to PBS2 in the path $\ket{0}_\text{R}$. 
For the reflected arm, after the transformation implemented by HWP3, the H-polarization component of~\eqref{hwptransform} leaves PBS2 in the path $\ket{1}_\text{R}$ 
and the V-polarization component of~\eqref{hwptransform} is reflected to the path $\ket{0}_\text{R}$.
The last stage of the channel is implemented by another half-wave plate (HWP5@45$^\circ$) introduced in the path $\ket{1}_\text{R}$. This device turns  $\ket{H}_{\text{S}} \rightarrow \ket{V}_{\text{S}}$. Thereby, the transformations implemented by this channel in the initial state $\ket{\psi_{\text{RC}}}$  can be written as the map (by using the notation $\ket{x} \ket{y} = \ket{x}_{\text{S}} \ket{y}_{\text{R}}$)
\begin{equation}
(\ket{H} - i\ket{V}) \ket{0} \rightarrow \ket{H}\ket{0} - i[\cos{2\theta_v}\ket{V}\ket{0} + \sin{2\theta_v} \ket{V} \ket{1}] \text{ , }
\label{optmap}
\end{equation}
up to a normalization factor $1/\sqrt{2}$ on both sides. If we consider the definition of the PD channel in terms of its Kraus operators~\cite{Nielsen:Book}, we obtain the map
\begin{align}
(\ket{0} - i\ket{1})\ket{0}  \rightarrow &\ket{0} \ket{0}  - i\left[ 1-p(\tau_{\text{d}}) \right]^{1/2} \ket{1} \ket{0} \nonumber \\&- i p^{1/2}(\tau_{\text{d}}) \ket{1} \ket{1}  \text{ , } \label{optmap2}
\end{align}
where $p(\tau_{\text{d}}) = 1 - e^{-\gamma \tau_{\text{d}}}$, being $\gamma$ the decay rate. By comparing Eqs.~(\ref{optmap}) and~(\ref{optmap2}), we get 
\begin{equation}
\cos^2 (2\theta_v) = 1-p(\tau_{\text{d}}) \text{ . }
\end{equation} 
Therefore, HWP3 simulates the time evolution during the PD channel. 
For the initial condition $p(\tau_{\text{d}}=0)=0$, where the system does not interact with the reservoir, we have $\theta_v = 0^{\circ}$. 
In this case, HWP3 does not implement any change in the polarization state and, as expected, nothing happens with the initial state. 
Consequently, coherence does not decrease. On the other hand, for the asymptotic behavior, $p(\tau_{\text{d}}\rightarrow \infty)=1$, HWP3 implements 
the maximum rotation in the polarization state and the state completely loses its coherence.

\subsection{Otto cycle with linear optical circuits}

In order to realize the Otto cycle, we start with the state preparation. As shown in~Fig~\ref{Fig1-2}, a vertically polarized DPSS laser ($1.5$~mW power, $\lambda = 532 nm$) passes through a quarter-wave plate QWP$1@-45^\circ$ to produce a right-circular polarization that is the analog of the initial state $\ket{\psi_{\text{RC}}}$. The laser beam passes into a spatial filter SF in order to improve the fundamental transverse mode quality. The initial state is verified by performing state tomography in the polarization of the laser beam at point $T_{\text{A}}$. Polarization tomography can be performed by following Ref.~\cite{Photonic-State-Tomography}. A detailed procedure to perform optical polarization tomography is presented in Appendix~\ref{tomopol}. 

The adiabatic expansion $AB$ is performed by the unitary evolution operator $U_{\text{e}}(\bar{\omega}\tau)$, where the gap expansion given by $\omega_{\text{e}}(t) = \omega_{0}(1 - t/\tau) + 2\omega_{0}(t/\tau)$ is realized by two half-wave plates HWP1 and HWP2, with their fast axes performing an angle of $\theta$ between them. As we show in Appendix~\ref{JonesMatrix}, by using the Jones matrices $S(\alpha)$ for polarization manipulation~\cite{Jones-i:41}, 
 it follows that the dimensionless quantity 
$\omega_{0} \tau$ can be associated with the angle $\theta$ between HWP1 and HWP2 through $(3/2)\omega_{0} \tau = \theta$. 
In our experiment, we consider $\theta = 3\pi/2$, so that $\omega\tau = \pi$. Note that, from the initial state, the circular polarization remains unchanged up to a global phase, which corresponds to the evolution of the eigenstates of the Hamiltonian. At point $T_\text{B}$, we perform the tomography of the evolved state in the same way as performed in $T_\text{A}$ (see Appendix~\ref{tomopol}).

Following the cycle, the step $BC$ corresponds to the hot reservoir. This part is simulated by PD1. Note that the amount of heat $Q_{\text{B} \rightarrow {C}}$ is related to the angle $\theta_V$, as described above to the evolution in the PD channel. A new tomography tracing out the environment is performed in $T_{\text{C}}$ in the two outputs of PD1 exactly as depicted in the measurement box at the end of the circuit.  The couples HWP$_{13}$/QWP$_2$ and HWP$_{14}$/QWP$_3$ are responsible by basis choices of tomography measurements while HWP$_{15}$ and the incidence of the two arms in PBS$_{TR}$ correspond to tracing out the environment as discussed in Ref.~\cite{PRA.78.Davidovich}.

The step $CD$ corresponds to the adiabatic compression and is also realized by two half-wave plates. Note that each output of PD1 passes through a couple of wave plates (HWP$_{6}$HWP$_{7}$ and HWP$_{8}$HWP$_{9}$) at the same angles of the couple HWP$_{1}$HWP$_{2}$ of step $AB$. This set of angle simulates the compression to the same initial volume, since the gap compression is considered as $\omega_{\text{c}}(t) = \omega_{0}(t/\tau) + \omega_{0}(1 - t/\tau)$. A new tomography of the state is performed at $T_{\text{D}}$. 

In order to complete the cycle, at step $DA$, it is necessary to return to the initial state. Therefore, the action of the PD channel should be undone. In order to accomplish such an assignment, we designed an optical circuit, namely, the inverted-PD channel (I-PD) to simulate the last step of the thermal engine (See the block I-PD in Fig.~\ref{Fig1-2}). Note that HWP$_{10}$ at $45^\circ$  undoes the action of HWP$_5$. In the interferometer, HWP$_{11}$ at $\theta_{V2} = \theta_V$ and HWP$_{12}$  at $\theta_{H2}=0$ undo the action of HWP$_3$  at $ + \theta_{V}$. PBS$_4$ regroups the arms $\ket{0}$ and $\ket{1}$, with the relative phase controlled by PZT$_2$ in order to obtain the initial state (circular polarization).
It is worth mentioning that our apparatus is not an actual thermalization process, but it is able to reproduce the desired output
thermal states at the end of the evolution. To this end, we used the standard procedure to simulate
decoherence in quantum circuits by taking the path degree of freedom as an ancilla, where the phase damping is recorded after the PD channel and recovered by the I-PD.  
These simulated thermal baths are fundamental in our implementation since we do not have a natural source
of thermal reservoir in our system. A tomography is performed at point $T_{A\prime}$ that corresponds to the initial state at $T_A$.

The thermodynamic quantities are evaluated by adopting regularly spaced $\theta_V$, namely 0$^\circ$, 8$^\circ$, 16$^\circ$, 22.5$^\circ$, 29$^\circ$, 37$^\circ$, and 45$^\circ$. The minimum angle corresponds to the absence of interaction between the qubit and the reservoir, namely, 
$\theta_V=0^\circ$ means that no heat is exchanged, while $\theta_V=45^\circ$ (maximum angle) means maximum heat exchange obtained for complete dephasing of the initial state.For each $\theta_V$, the whole cycle is performed so that hot and cold reservoirs with different temperature ratios are simulated.

\vspace{0.3cm}

\section{Results and discussion}

\begin{figure}[t!]
	\centering
	\includegraphics[scale=0.35,clip,trim=0mm 0mm 0mm 0mm]{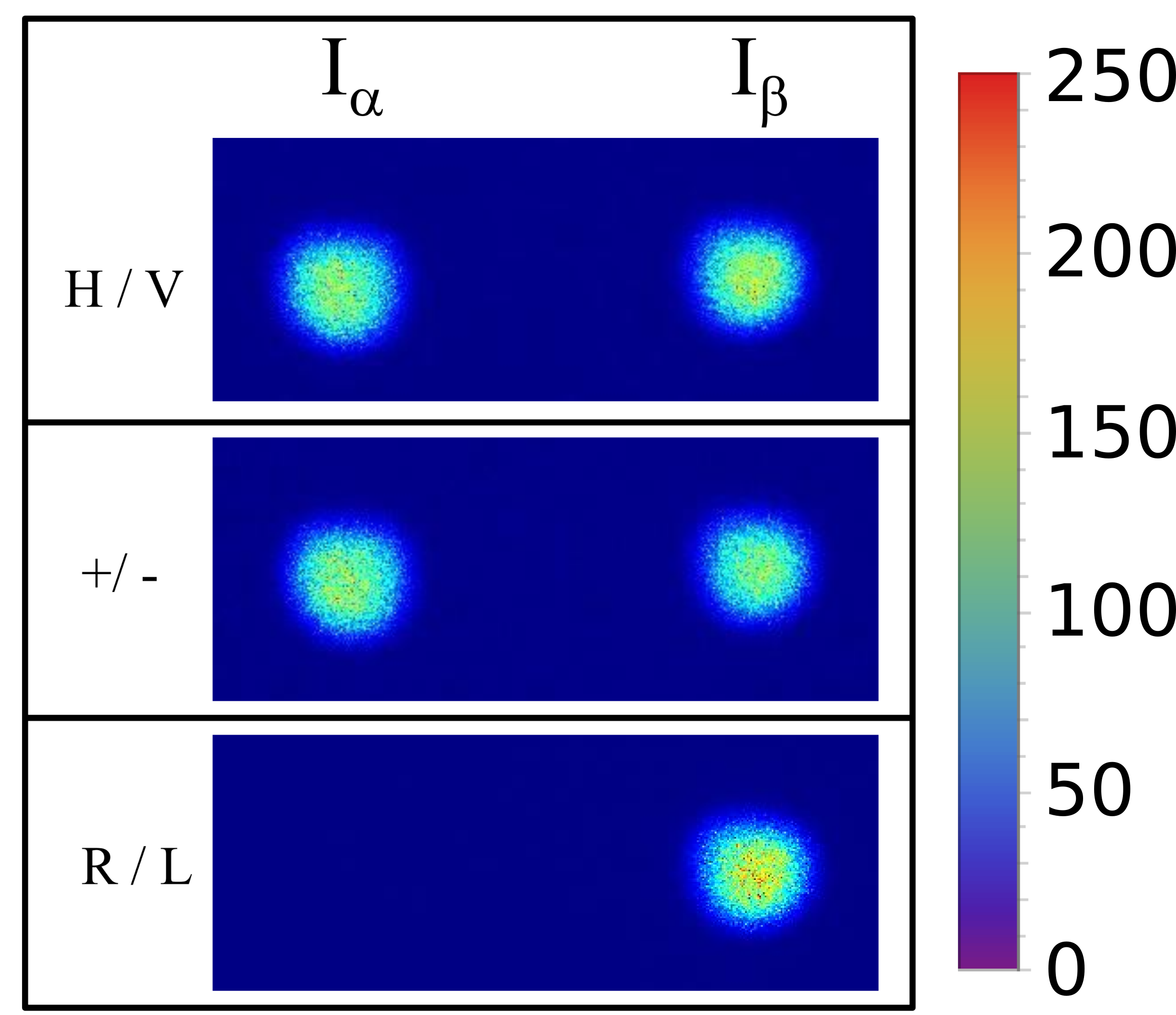}
	\caption{ Tomographic images for the initial state $\rho_i$ with intensities $I_{\alpha}$ and $I_{\beta}$ for the basis $\{\alpha, \beta$\} = \{H,V\}, \{D,AD\}, and \{R,L\}. The normalized intensities were used to obtain the density matrix $\rho_{\text{ini}}^{\text{exp}}$. Color bar represents the gray level intensity of the digital images.}
	\label{imgtomo}
\end{figure}

\begin{figure*}[ht!]
	\centering
	\subfloat[Internal energy variation $\Delta U$]{\includegraphics[scale=0.41]{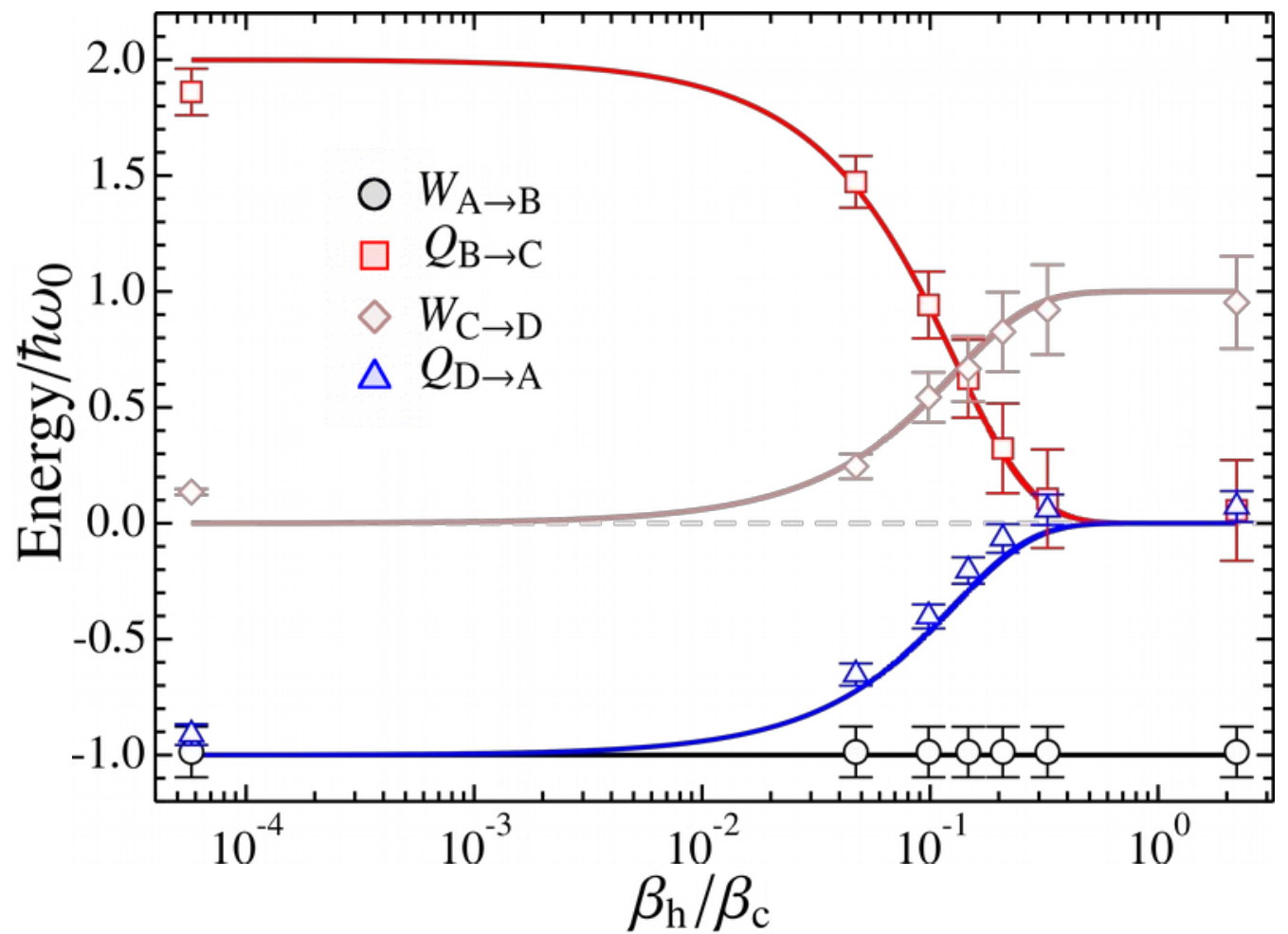}\label{Fig2a}} \quad \subfloat[Extracted work and energy balance]{\includegraphics[scale=0.41]{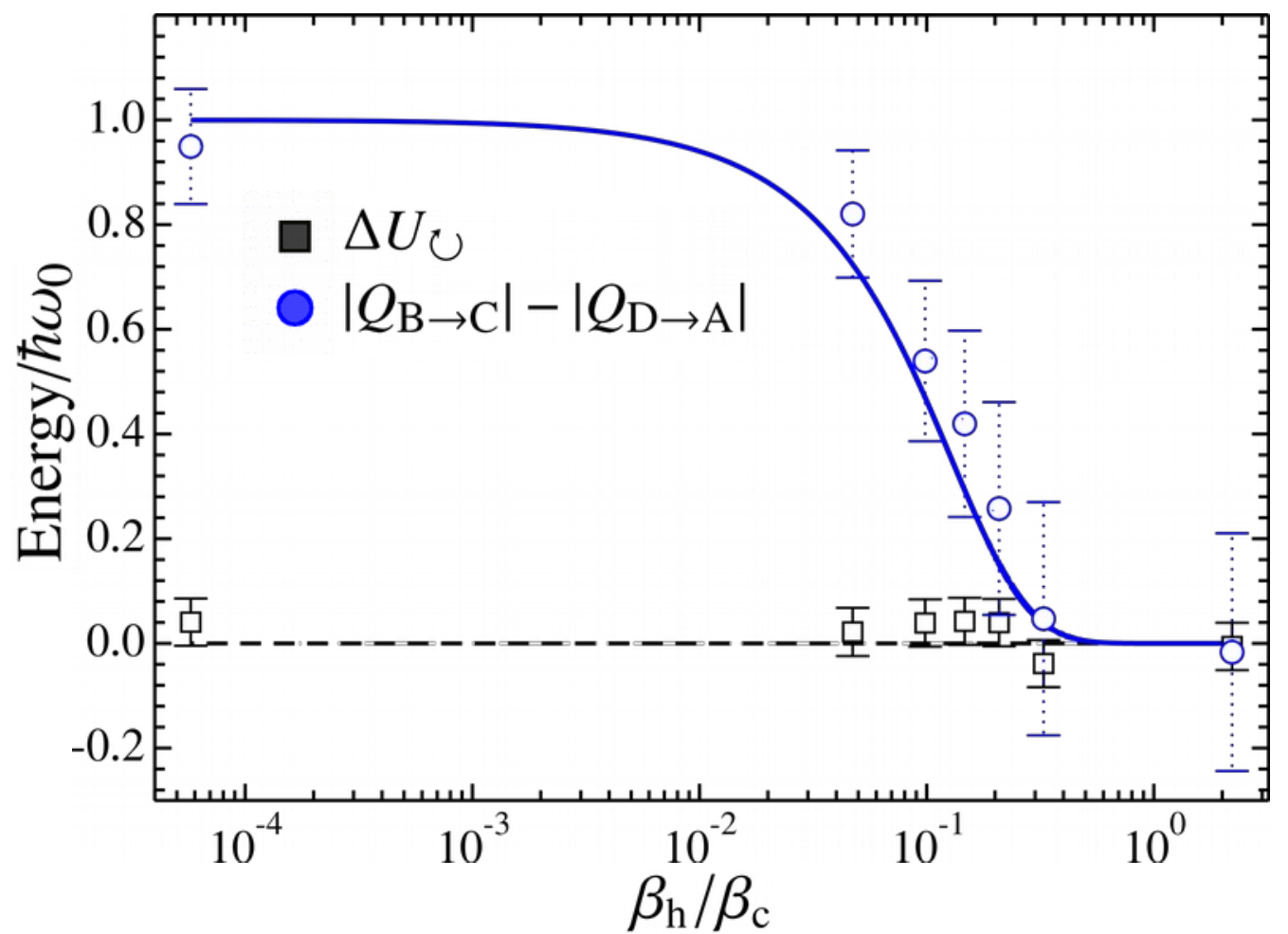}\label{Fig2b}} \quad \subfloat[Entropy production $\langle \Sigma (\beta_{\text{h}}/\beta_{\text{c}}) \rangle$]{\includegraphics[scale=0.41]{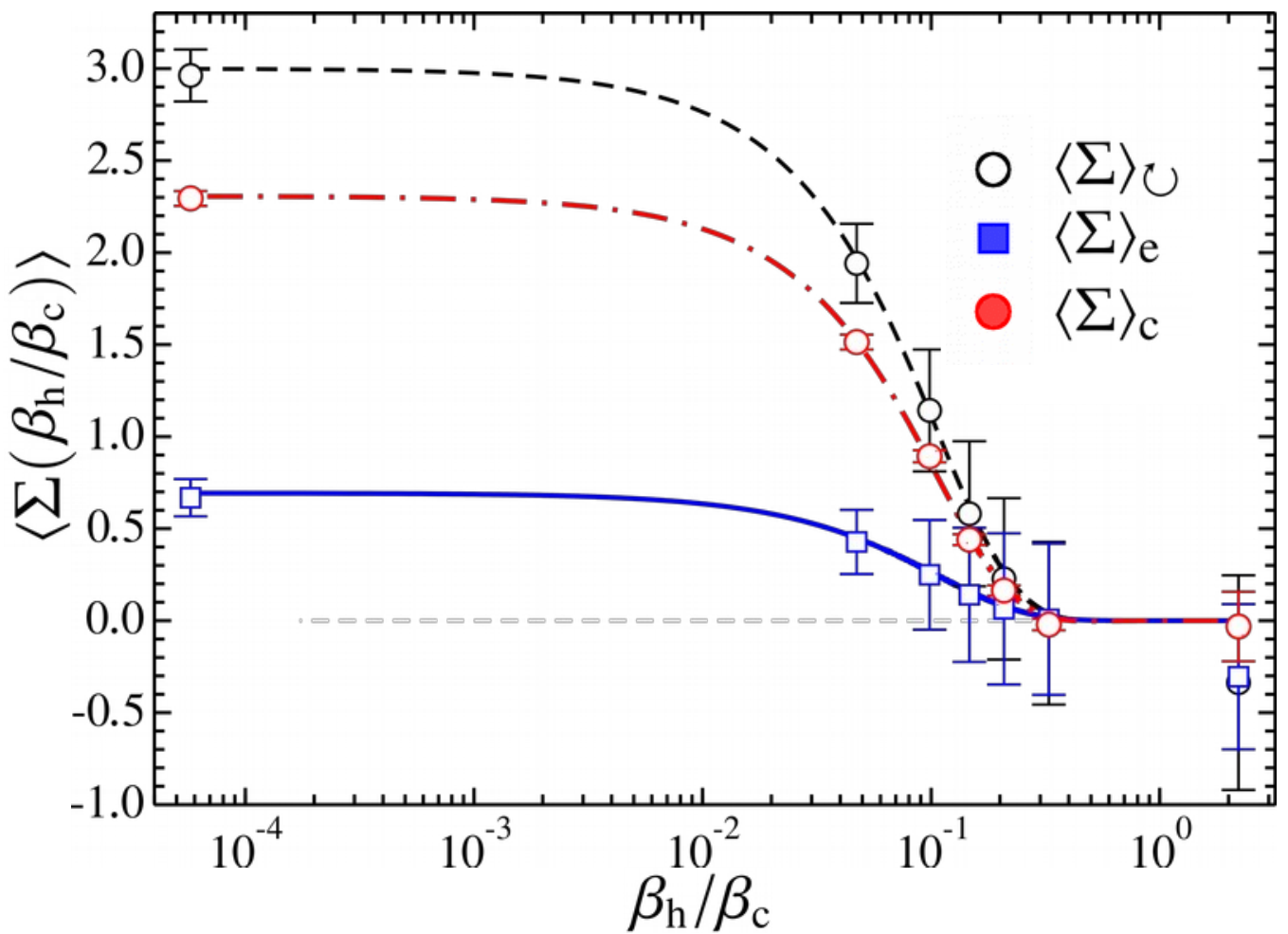}\label{Fig2c}}
	\caption{(\ref{Fig2a}) Heat $Q$ and work $W$ at each step of the Otto cycle in unities of $\hbar \omega_{0}$ as functions of $\beta_{\text{h}}/\beta_{\text{c}}$. (\ref{Fig2b}) Energy balance $\Delta U_{\circlearrowright}$  and extracted work $|Q_{\text{B}\rightarrow\text{C}}| - |Q_{\text{D}\rightarrow\text{A}}|$ for a closed cycle. (\ref{Fig2c}) Entropy production $\langle \Sigma \rangle_{\text{e}}$ and $\langle \Sigma \rangle_{\text{c}}$ associated with the thermalization process after the  unitary expansion and compression steps, respectively, as well as the entropy $\langle \Sigma \rangle_{\circlearrowright}$ yielded in a closed cycle. The plots (\ref{Fig2a}--\ref{Fig2c}) represent the expected experimental results we  would obtain in a  genuine implementation of a thermal machine such that the working substance is initialized in the thermal state of $H_{\text{e}}(0)$ at temperature $k_{\text{B}}T_{\text{c}} \approx \hbar \omega /3$, where $k_{\text{B}}$ is the Boltzmann constant. Continuum lines are theoretical predictions computed from Eqs.~(\ref{Wab}--\ref{Qda}), while the points represent their respective experimental data computed from Eqs.~(\ref{workDM}) and~(\ref{heatDM}). The error bars have been obtained by the error propagation of 
	the intensity uncertainty in the experimental density matrix elements and in Eqs. (\ref{workDM}), (\ref{heatDM}), and (\ref{ApTradS}), respectively.}
	\label{Fig2}
\end{figure*}
Let us start with the characterization of the initial state (right circular polarization). After preparation, we perform a tomographic measurement according to the procedure detailed in Appendix A. The images associated to each measured basis are presented in Fig.~\ref{imgtomo}. Note that for the R/L basis only one port (I$_{\beta}$) is lightened, as expected. With the intensities I$_{\alpha}$ and I$_{\beta}$ we can  reconstruct the density matrix analog to that of the quantum state $\ket{\psi_{\text{RC}}}$. The density matrix for the initial state $\rho_{\text{ini}}^{\text{exp}}$ is
\begin{equation}
\label{rhoiexp}
    \rho_{\text{ini}}^{\text{exp}} = \left ( \begin{array}{cc}
0.5134 & 0.0033 + 0.4999\;i \\
0.0033 - 0.4999\;i & 0.4865\\
\end{array} \right ) \text{ . }
\end{equation}
This result is in very good agreement with the theoretical prediction, which is given by
\begin{equation}
\label{rhoi}
    \rho_{\text{ini}} = \left ( \begin{array}{cc}
0.5 & 0.5i \\
-0.5i & 0.5 \\
\end{array} \right ) \text{ . }
\end{equation}
From Eqs.~(\ref{Wab}--\ref{Qda}) we compute heat $Q$ and work $W$ from the internal energy variation $\Delta U$ at each step of the Otto cycle.  
By using the definition of internal energy as $U = \tr{H\rho}$, for some reference Hamiltonian $H$, we evaluate $U$ from the experimental density matrix 
after each thermodynamical process, which is obtained by performing quantum state tomography. Then, we have
\begin{eqnarray}
W &=& \tr{\rho(\tau) H(\tau) - \rho(0) H(0)} \text{ , } \label{workDM} \\  
Q &=& \tr{H(\tau)[\rho(0) - \rho(\tau)]} \text{ . } \label{heatDM}
\end{eqnarray}
where $W$ is the work performed during a unitary evolution and $Q$ is obtained through a nonunitary process keeping the Hamiltonian constant~(See Appendix~\ref{ApHeatWork}). In our experimental implementation, we start from the thermal state of the Hamiltonian $H_{\text{e}}(0) = \hbar\omega_{0} \sigma_{y}$, with $\tanh (\hbar \omega_{0} \beta_{\text{c}}) \approx 1$ (corresponding to $\hbar \omega_{0} \beta_{\text{c}} \approx 3$). In this case, the initial state is $\rho_{\text{ini}}^{\text{th}} \approx \ket{\psi_{\text{RC}}}\bra{\psi_{\text{RC}}}$. The results are presented in Fig~\ref{Fig2}. In Fig~\ref{Fig2a}, we show the  internal energy variation $\Delta U/\hbar \omega_0$ as a function of $\hbar \omega_0 \beta_h$ for each step of the cycle for seven different values of the inverse hot temperature $\beta_h$. In this plot, work and heat have experimentally been obtained by Eqs.~(\ref{workDM}) and~(\ref{heatDM}), respectively. For a closed cycle, $\Delta U $ must be zero, which can be observed by summing up $W$ and $Q$ for all the curves at a 
fixed value of $\beta_h$. Fig.~\ref{Fig2b} shows the extracted work, quantified from difference $|Q_{\text{B}\rightarrow\text{C}}|-|Q_{\text{D}\rightarrow\text{A}}|$, due to the coupling of the system with thermal baths at different inverse temperatures $\beta_{\text{c}}$ and $\beta_{\text{h}}$. As expected, the extracted work decreases as the inverse hot temperature 
$\beta_h$ increases. In addition, note that the energy balance $\Delta U_{\circlearrowright}$, is kept close to zero, as theoretically predicted.

To study the entropy production in the thermodynamic cycle, we consider the analysis of the expansion/compression step followed by the 
subsequent thermalization process. The irreversible contribution to the entropy variation is then given by~\cite{Procaccia:76,Jarzynski:97,Parrondo:09,Lindblad:Book}
\begin{eqnarray}
\langle \Sigma \rangle_{\text{e/c}} = \Delta S_{\text{e/c}} - \beta_{\text{h/c}}Q_{\text{e/c}}\text{ , } \label{ApTradS}
\end{eqnarray}
where $\Delta S_{\text{e/c}} = S^{\text{fin}}_{\text{e/c}} - S^{\text{ini}}_{\text{e/c}}$ is the (von Neumann) entropy variation during the thermalization process after expansion/compression step, with $S_{\text{e/c}} = - \tr{\rho^{\text{hot/cold}}_{\text{th}} \ln \rho^{\text{hot/cold}}_{\text{th}} }$, and $Q_{\text{e/c}}$ is the amount of heat exchanged during such a  process. It is possible to show that the above equation can be written as the quantum relative entropy (Kullback–Leibler divergence) $\langle \Sigma \rangle_{\text{e/c}} = \Dcal[\rho_{\text{e/c}}(\tau)||\rho^{\text{hot/cold}}_{\text{th}}] =  \rho_{\text{e/c}}(\tau) \ln \rho_{\text{e/c}}(\tau) - \rho_{\text{e/c}}(\tau) \ln \rho^{\text{hot/cold}}_{\text{th}}$, where $\rho_{\text{e/c}}(\tau)$ is the state after expansion/compression step and $\rho^{\text{hot/cold}}_{\text{th}}$ is the 
thermal state at inverse temperature $\beta_{\text{h/c}}$ to be achieved as a subsequent thermalization process~\cite{Lindblad:Book}~(see Appendix~\ref{Entropy} for more details). As originally proposed, $\langle \Sigma \rangle_{\text{e/c}}$ quantifies the lag between the non-equilibrium state after the expansion/compression unitary step and the 
target equilibrium thermal state. For a recent experimental implementation in NMR, see Ref.~\cite{Batalhao:15}. For thermodynamical cycles, the entropy production accounts for the dissipated energy during the expansion and compression steps, which may quantify 
\textit{quantum friction} during the quantum evolution~\cite{Adolfo:14,Plastina:14}. 
The results are shown in Fig.~\ref{Fig2c}. Observe that the individual amounts of entropy production $\langle \Sigma \rangle_{\text{e}}$ and $\langle \Sigma \rangle_{\text{c}}$ associated with the thermalization after the (unitary) expansion and compression steps, respectively, are nonvanishing, which implies a nonvanishing total entropy production $\langle \Sigma \rangle_{\circlearrowright}$  for thermal 
baths with distinct inverse temperatures $\beta_{\text{c}}$ and $\beta_{\text{h}}$. Notice that, as theoretically predicted, $\langle \Sigma \rangle_{\circlearrowright}$ vanishes as $\beta_{\text{h}}$ gets nearer $\beta_{\text{c}}$. The experimental images and the reconstructed density matrix for each step  considering $\theta_V=22.5^\circ$ are presented in Appendix~\ref{imagmatrix}.  Note that the first experimental points in Fig.\ref{Fig2a}), \ref{Fig2b}) and \ref{Fig2c}) correspond to $\theta_V=45^\circ$ ($\beta_h / \beta_c < 10^{-4} $) and the last ones corresponds to $\theta_V=0^\circ$ ($\beta_h / \beta_c = 1$).

\section{Conclusions}

In summary, we introduced a map from a single-qubit thermal machine into a single-photon setup. The feasibility of this proposal has been experimentally tested by an all-optical 
experiment realized through an intense laser beam. By using the polarization degree of freedom of the laser beam, we encoded a qubit as the working substance, while the two thermal baths are simulated by  correlating the polarization with an auxiliary degree of freedom, which was the propagation path in our experiment. We have then shown how different thermal baths can be implemented with optical devices, with the difference of temperatures controllable through a dimensionless parameter associated with a combination of half-wave plates. 
Agreement between experimental and theoretical results is remarkable, with errors within a $5\%$ range.  It is worth emphasizing that we are proposing  
an all-optical simulation of a thermal machine, which aims at reproducing analogues of Gibbs states, heat transfers and 
work extraction.  However, our investigation opens perspectives for 
optical implementations of other protocols in quantum thermodynamics. In this scenario, the experimental discussion of the performance of quantum refrigerators~\cite{Lutz:16,Noah:10,Brask:15} with optical devices and optical quantum thermometers~\cite{Haupt:14,Hofer:17} is left as future research.

\section*{Acknowledgments}
The authors acknowledge financial support from the Brazilian funding agencies Conselho Nacional de Desenvolvimento Cient\'{\i}fico e Tecnol\'ogico (CNPq), 
Funda\c{c}\~ao Carlos Chagas Filho de Amparo \`a Pesquisa do Estado do Rio de Janeiro (FAPERJ), 
Coordena\c{c}\~ao de Aperfei\c{c}oamento de Pessoal de N\'{\i}vel Superior (CAPES) (Finance Code 001), 
and the Brazilian National Institute for Science and Technology of Quantum Information (INCT-IQ).

\appendix

\section{Heat and work in thermodynamic processes driven by time-local master equations} \label{ApHeatWork}

Let us consider a quantum system evolving driven by a time-dependent Hamiltonian $H(t)$ and coupled to a reservoir governed by a time-local master equation
\begin{align}
\dot{\rho}(t) = \frac{1}{i\hbar} [H(t),\rho(t)] + \Rcal[\rho(t)] \text{ , } \label{ApDynLind}
\end{align}
where $\Rcal[\bullet]$ describes the coupling between the environment and the system, so that we can recover a free-decohering evolution as $\Rcal[\rho(t)]=0$ for all $t\in[0,\tau_{\text{ev}}]$, with $\tau_{\text{ev}}$ being the total evolution time. The internal energy of the system can be computed from $U(t) = \tr{\rho(t)H(t)}$, 
yielding
\begin{align}
dU(t) &= \frac{d}{dt} \left[\tr{\rho(t)H(t)}\right]dt \nonumber\\&= \tr{\dot{\rho}(t)H(t)}dt + \tr{\rho(t)\dot{H}(t)}dt \text{ . } \label{ApdU}
\end{align}
Now, let us consider two distinct situations. First, consider that the system evolves under the action of a time-independent Hamiltonian. Thus, we get
\begin{align}
dU(t)|_{H(t)=H} = \tr{\dot{\rho}(t)H}dt \text{ . } \label{AquiIn}
\end{align}
We can then use the Eq.~\eqref{ApDynLind} and write
\begin{align}
dU(t)|_{H(t)=H} &= \tr{\left[\frac{1}{i\hbar} [H,\rho(t)] + \Rcal[\rho(t)]\right]H}dt \nonumber\\
	  &= \frac{1}{i\hbar}\tr{[H,\rho(t)]H}dt + \tr{\Rcal[\rho(t)]H}dt \text{ . }
\end{align}
By using the cyclic property of the trace, we obtain $\tr{[H,\rho(t)]H} = 0$, so that
\begin{align}
dU(t)|_{H(t)=H} &= \tr{\Rcal[\rho(t)]H}dt\text{ . }
\end{align}
This means that any internal energy variation will be due to fluctuations in the energy level populations arising from the coupling of the system with its environment. 
For this reason, we identify this change in the internal energy as \textit{heat} and we write
\begin{align}
dU(t)|_{H(t)=H} &= dQ = \tr{\Rcal[\rho(t)]H}dt\text{ . }
\end{align}
Thus, the total exchanged heat is obtained by integration of the above equation as
\begin{align}
Q = \int_{\tau_0}^{\tau}dQ = \int_{\tau_0}^{\tau} dU(t)|_{H(t)=H} \text{ . }
\end{align}
By using that $\int_{t_{1}}^{t_{2}}df(t) = f(t_{2}) - f(t_{1})$ for any analytical function $f$ in interval $\Ical = [t_{1},t_{2}]$, we get
\begin{align}
Q = \int_{\tau_0}^{\tau} dU(t)|_{H(t)=H} = \tr{H[\rho(\tau_0) - \rho(\tau)]} \text{ . }
\end{align}
On the other hand, consider the case where we have a time-dependent Hamiltonian, but the system is decoupled from its reservoir [$\Rcal[\rho(t)]=0$]. 
Thus, from Eq.~\eqref{ApdU} we get
\begin{align}
dU(t)|_{\Rcal[\rho(t)]=0} &= \tr{\dot{\rho}(t)H(t)}dt + \tr{\rho(t)\dot{H}(t)}dt \text{ , } 
\end{align}
where we can use the Eq.~\eqref{ApDynLind} to write
\begin{align}
\tr{\dot{\rho}(t)H(t)} &= \frac{1}{i\hbar}\tr{ [H(t),\rho(t)] H(t)} \nonumber \\
&= \frac{1}{i\hbar}\left[ \tr{ H(t)\rho(t)H(t)} - \tr{ \rho(t)H(t)H(t)}\right] \nonumber \\
&= \frac{1}{i\hbar}\left[ \tr{ \rho(t)H(t)H(t)} - \tr{ \rho(t)H(t)H(t)}\right] = 0 \text{. }
\end{align}
Therefore
\begin{align}
dU(t)|_{\Rcal[\rho(t)]=0} &= \tr{\rho(t)\dot{H}(t)}dt \text{ . } 
\end{align}
This means that, when we have a unitary evolution, changes in the internal energy may be achieved by variations in the instantaneous energy eigenspectrum of the time-dependent Hamiltonian. So, by taking the Hamiltonian as controlled by external fields governing the gap expansion/compression, we identify $dU(t)|_{\Rcal[\rho(t)]=0}$ as an amount 
of work performed by/on these fields and we write
\begin{align}
dU(t)|_{\Rcal[\rho(t)]=0} &= dW(t) = \tr{\rho(t)\dot{H}(t)}dt \text{ . } 
\end{align}
Hence, the work performed on/by the system can be obtained by integration of the above equation as
\begin{align}
W = \int_{\tau_0}^{\tau}dW(t) = \int_{\tau_0}^{\tau}dU(t)|_{\Rcal[\rho(t)]=0} \text{ , } 
\end{align}
and therefore we conclude that
\begin{align}
W = \Delta U(\tau) = \tr{\rho(\tau) H(\tau) - \rho(\tau_0) H(\tau_0)} \text{ . } 
\end{align}

\section{Optical polarization tomography}\label{tomopol}
	
The state of one qubit can be described in terms of the density operator 
\begin{equation}
    \rho = \frac{1}{2}\left(\1 + \sum_{i=1}^3\;r_i\;\sigma_i\right) \text{ , }
    \label{rhoA}
\end{equation}
where the matrices ($\sigma_1$,$\sigma_2$,$\sigma_3$) are the Pauli matrices and ($r_1$,$r_2$,$r_3$) are the components of the Bloch vector. 
Considering optical polarization, it is possible  to rewrite the density operator given by Eq. (\ref{rhoA}) in terms of the Stokes parameters 
$S_i$ ($i$=0,1,2,3)~\cite{Photonic-State-Tomography} so that we have
\begin{equation}
    \rho = \frac{1}{2}\left(S_0 \1 + \sum_{i=1}^3\;S_i\;\sigma_i\right) \text{ , }
    \label{rhoS}
\end{equation}
where $S_0 = \tr{\1\rho}=1$ and each $S_i$ ($i$=0,1,2,3) is defined by
\begin{equation}
    S_i = \tr{\sigma_i\rho} \text{ . }
\end{equation}

As provided in Ref.~\cite{Photonic-State-Tomography}, in optical polarization tomography, 
each component $S_i$ can be obtained by defining a set of projective measurements in the polarization states of light as 
\begin{eqnarray}
S_0 &=& P_{|H \rangle} + P_{|V \rangle} \text{ , } \label{S_0} \\ 
S_1 &=& P_{|D \rangle} - P_{|AD \rangle} \text{ , } \label{S_1}\\ 
S_2 &=& P_{|L \rangle} - P_{|R \rangle} \text{ , } \label{S_2} \\ 
S_3 &=& P_{|H \rangle} - P_{|V \rangle} \text{ , } \label{S_3}
\end{eqnarray}
where $|H \rangle$ ($|V \rangle$) represents the horizontal (vertical) polarization, $|D \rangle$ ($|AD \rangle$) represents the linear diagonal (linear anti-diagonal) polarization,  
$|L \rangle$ ($|R \rangle$) represents the left-circular (right-circular) polarization of light, and $P_{\ket{\alpha}}$ is the probability of obtain the component 
$\ket{\alpha}$, with $\alpha= \{H,V\}, \{D, AD\}, \{R, L\}$.

\begin{figure}[t!]
	\centering
	\includegraphics[scale=1.0]{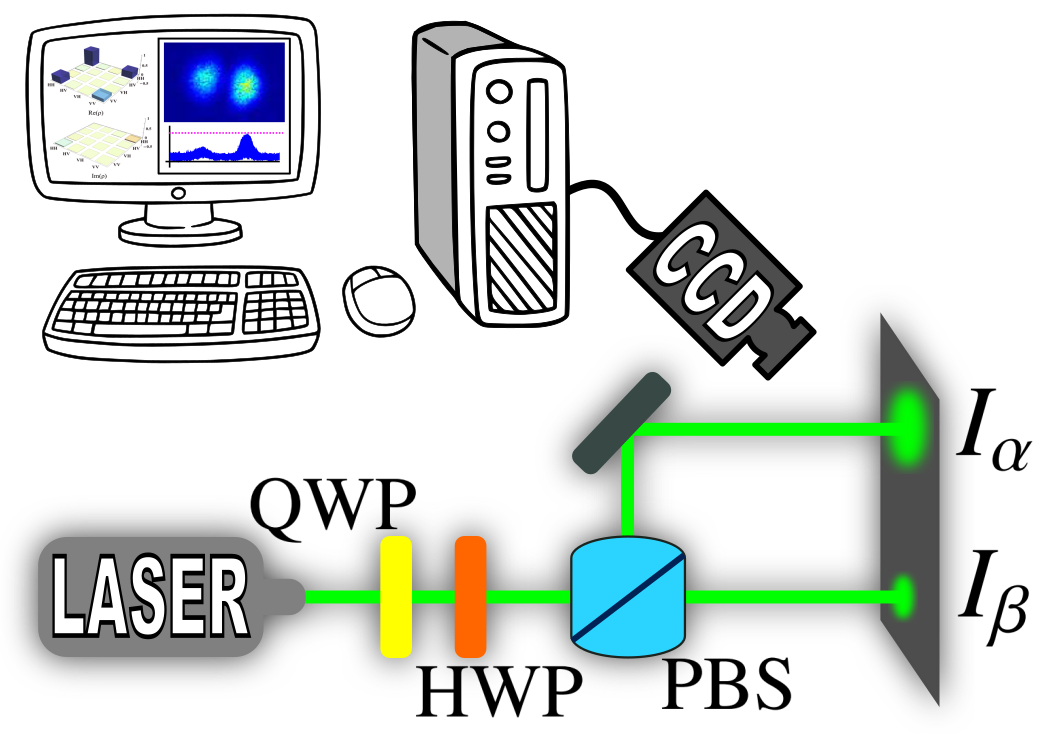}
	\caption{Experimental circuit to implement the optical polarization tomography.}
	\label{expTomo}
\end{figure}

Fig.~\ref{expTomo}  shows how we can implement the polarization tomography. Experimentally, to obtain the Stokes parameters $S_0$ and $S_3$, we make a projective measurement in the basis $\{H, V\}$ using a Polarization Beam Splitter (PBS). To obtain the Stokes parameter $S_1$ the projective measure is performed by using the HWP@22.5$^{\circ}$ in association with the PBS, that corresponds to measurement in the basis $\{D, AD\}$. Finally, to obtain the Stokes parameters $S_2$ we use the QWP@0$^{\circ}$, the HWP@22.5$^{\circ}$, and the PBS in order to measurement in the basis $\{R,L\}$. Lastly, a charged-coupled device (CCD) records one single image with the intensity of each component projected on a screen.

Considering our apparatus, we measure simultaneously the intensity of each projected components $I_{\alpha}$ and $I_{\beta}$, 
being $\beta$ the complementary (orthogonal) basis component to $\alpha$, e.g., for $\alpha=H$, then $\beta=V$. 
We can associate the probability $P_{\ket{\alpha}}$ with the normalized intensity as
\begin{equation}
    P_{\ket{\alpha}} = I_{\ket{\alpha}}= \frac{I_{\alpha}}{I_{\alpha} + I_{\beta}}. 
\end{equation}
Then, the Stokes parameters $S_i$ can be obtained from the intensities of tomographic measurements and we can reconstruct the density matrix for any polarization state by using Eq.~\eqref{rhoS}.

\section{Simulating expansion/compression step by Jones matrix} \label{JonesMatrix}

The polarization rotation Jones matrix is defined by
\begin{eqnarray}
S(\alpha) = \begin{bmatrix}
\cos (\alpha) & - \sin (\alpha) \\
\sin (\alpha) &  \cos (\alpha)
\end{bmatrix} \text{ . } \label{EqJonesMatrix}
\end{eqnarray}
 Eq.~(\ref{EqJonesMatrix}) may be used to encode changes in character of light as it passes through a partial polarizer, for example, with $\alpha$ denoting the rotation angle between the initial and final polarizations. The matrix $S(\alpha)$ is in the SO$(2)$ group. Then, it can be used to provide a transformation of a vector $\ket{\psi}$ to a transformed vector $\ket{\psi^{\prime}}$ such that the norm of $\ket{\psi}$ is preserved. We can write $S(\alpha) = e^{-i J \alpha} $, with the generator $J$ given by~\cite{Tung:Book}
\begin{eqnarray}
J = \begin{bmatrix}
0 & - i \\
i &  0
\end{bmatrix} \text{ , }
\end{eqnarray}

On the other hand, the evolution operator of the working substance driven by Hamiltonian $H = \hbar \omega(t)\sigma_{y}$ is given by
\begin{eqnarray}
U(t,t_{0}) = \exp \left[ -\frac{i}{\hbar} \int_{t_{0}}^{t} H(\xi) d \xi \right] \text{ , }
\end{eqnarray}
which can be rewritten as
\begin{eqnarray}
U(t,t_{0}) = \exp \left[ - i \bar{\omega} (t-t_{0}) \sigma_{y} \right] \text{ , }
\end{eqnarray}
where $ \bar{\omega} = (t-t_{0})^{-1}\int_{t_{0}}^{t} \omega(\xi) d \xi$. In this case, $\sigma_{y}$ can be viewed as the dynamical generator associated with 
the evolution operator $U(t,t_{0})$, 
being exactly the same as the generator of $J$. Thus, we conclude that there is a correspondence between evolution operator $U(t,t_{0})$ and the Jones matrix $S(\alpha)$ by taking
\begin{eqnarray}
\alpha = \bar{\omega} (t-t_{0}) = \int_{t_{0}}^{t} \omega(\xi) d \xi \text{ . }
\end{eqnarray}
\begin{figure*}[t!]
	\centering
	\includegraphics[scale=0.13]{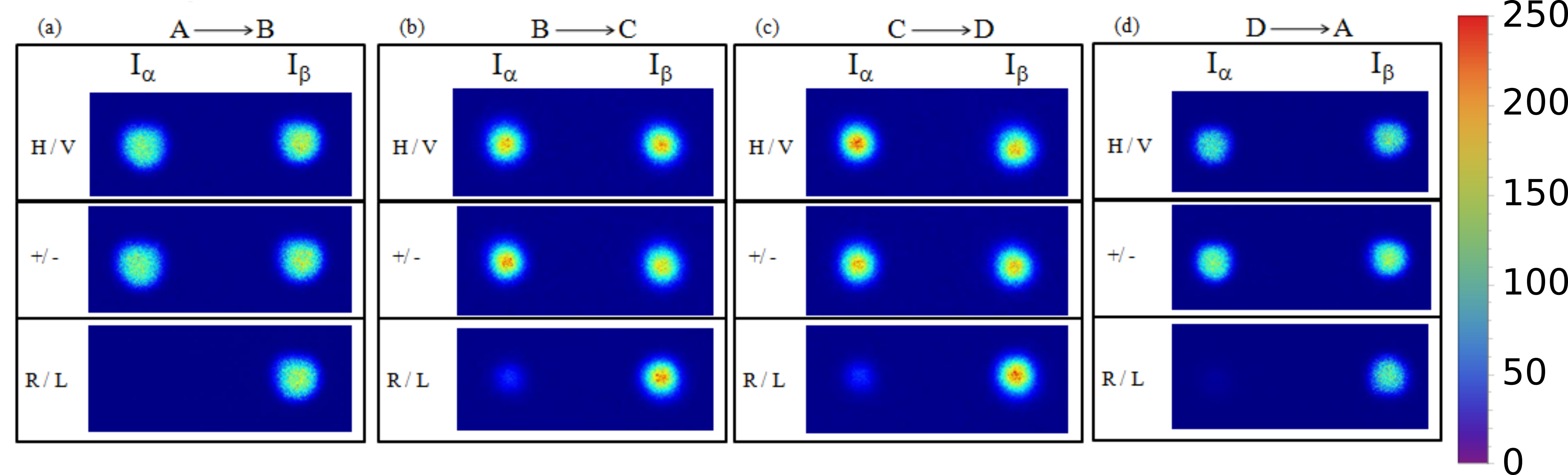}
	\caption{Tomographic images at the end of each step of the cycle with intensities $I_{\alpha}$ and $I_{\beta}$ for the basis $\{\alpha, \beta$\} = \{H,V\}, \{D,AD\}, and \{R,L\}. The normalized intensities were used to obtain the density matrix of each step of the cycle. Color bar represents the gray level intensity of the digital images.}
	\label{modes}
\end{figure*}
Therefore, by setting the expansion/compression duration as $t-t_{0} = \tau$ and the frequency $\omega$, it is possible to adjust $\alpha$ to simulate the desired dynamics. 
In particular, in the expansion step considered in main text we have $\omega_{\text{e}}(t) = \omega_{0}(1 - t/\tau) + n\omega_{0}(t/\tau)$ (with $t_{0}=0$), where $n>1$ in 
order to obtain $\omega_{\text{e}}(0)<\omega_{\text{e}}(\tau)$. Then, we get the {\it expansion Jones parameter} $\alpha_{\text{e}}$ as
\begin{eqnarray}
\alpha_{\text{e}} = \int_{0}^{\tau} \omega_{\text{e}}(\xi) d \xi = \frac{ (n+1) \omega_{0} \tau}{2} \text{ . }
\end{eqnarray}
On the other hand, in the compression step, we need to get $\omega_{\text{c}}(0)>\omega_{\text{c}}(\tau)$, so that we write $\omega_{\text{c}}(t) = n\omega_{0}(1 - t/\tau) + \omega_{0}(t/\tau)$. Therefore the {\it compression Jones parameter} $\alpha_{\text{c}}$ reads
\begin{eqnarray}
\alpha_{\text{c}} = \int_{0}^{\tau} \omega_{\text{c}}(\xi) d \xi = \frac{ (n+1) \omega_{0} \tau}{2} \text{ . }
\end{eqnarray}
This means that we can simulate the expansion and compression steps with the same Jones parameter. In an experimental approach, the Jones matrix in Eq.~\eqref{EqJonesMatrix} can be implemented by using an arrangement of two wave plates HWP$_{\theta}$, where the parameter $\theta$ is associated with the angle between the fast axis and the vertical direction. The matrix representation of the HWP is
\begin{eqnarray}
\text{HWP}_{\theta} = \begin{bmatrix} \cos\theta & \sin\theta \\ \sin\theta & -\cos\theta
\end{bmatrix} \text{ , }
\end{eqnarray}
so that we can obtain the following result
\begin{eqnarray}
S(\alpha) = \text{HWP}_{2\alpha} \cdot \text{HWP}_{\alpha} = \begin{bmatrix} \cos (\alpha) & - \sin (\alpha) \\ \sin (\alpha) & \cos (\alpha)
\end{bmatrix} \text{ . }
\end{eqnarray}
In conclusion, the HWP angle needs to be adjusted as $\alpha = 3 \tau \omega_{0}/2$, so that the first HWP is $3 \tau \omega_{0}/2$ and the second one is $3 \tau \omega_{0}$, 
implying a relative angle between the HWPs given by $\theta_{\text{r}} = 3 \tau \omega_{0}/2$.

\section{Experimental images and density matrix for a complete cycle}\label{imagmatrix}

In this Appendix, we will present the resulting density matrix for each step of the engine for $\theta_V=22.5^{\circ}$, in order to show a complete measurement set of the experiment. Fig.~\ref{modes} presents the images obtained from tomographic measurement for each step of the cycle. 


Regarding the Otto's cycle in Fig.~\ref{Fig1-2}, the images of step AB are presented in Fig.~\ref{modes}-(a). 
The theoretical values of Stokes parameters for this case are $S_1=0$, $S_2$=1, and $S_3=0$. Note that only  R/L basis measurement should have a value different from zero ($I_{\alpha} \neq I_{\beta}$), which can be inferred from Fig.~\ref{modes}-(a). Then, the experimental density matrix for the step AB is given by
\begin{equation}
    \rho_{A\rightarrow B}^{\text{exp}} = \left ( \begin{array}{cc}
0.49663 & 0.0045 + 0.4966\;i \\
0.0045 - 0.4966\;i & 0.5033\\
\end{array} \right ) \text{ . }
\end{equation}
The resulting images for the end of the step BC are shown in Fig.~\ref{modes}-(b). Here, the expected theoretical values for the Stokes parameters are 
$S_1 = 0$, $S_2 =-0.7071$, and $S_3 = 0$, that means $I_{\alpha}\neq 0$, as observed in Fig.~\ref{modes}-(b). The resulting density matrix is given by
\begin{equation}
    \rho_{B\rightarrow C}^{\text{exp}} = \left ( \begin{array}{cc}
0.5057 & 0.0174 + 0.3407\;i \\
0.0174 - 0.3407\;i & 0.4942\\
\end{array} \right ) \text{ . }
\end{equation}
Note the change of the state due to the PD channel at this stage. 
Following the cycle, the images for the end of the step  CD are shown in Fig.~\ref{modes}-(c). The obtained density matrix associated with the stage is
\begin{equation}
    \rho_{C\rightarrow D}^{\text{exp}} = \left ( \begin{array}{cc}
0.5190 & 0.0041 + 0.3482\;i \\
0.0041 - 0.3482\;i & 0.4809\\
\end{array} \right ) \text{ . }
\end{equation}
As can be noted $\rho_{C\rightarrow D}^{\text{exp}}$ is very similar to $\rho_{B\rightarrow C}^{\text{exp}}$, as expected, since the performed unitary operation does not change the polarization. 
The last step of the Otto's cycle (DA) gives us the images presented in Fig.~\ref{modes}-(d). The respective density matrix is given by
\begin{equation}
    \rho_{D\rightarrow A}^{\text{exp}} = \left ( \begin{array}{cc}
0.5118 & 0.0115 + 0.4503\;i \\
0.0115 - 0.4503\;i & 0.4881\\
\end{array} \right ) \text{ . }
\end{equation}
Note that, if we compare the experimental density matrices $\rho_{\text{ini}}^{\text{exp}}$ and $\rho_{D\rightarrow A}^{\text{exp}}$ we can see that they are approximately equal, with the small differences due to the losses after the laser beam passes through all the optical elements of the experiment.

\section{Entropy production and quantum relative entropy} \label{Entropy}

In order to study the relation between entropy production and quantum relative entropy, we will follow a similar procedure as preformed in 
Refs.~\cite{Deffner:10,Apollaro:15,Plastina:14}. This approach has also been previously discussed in Ref.~\cite{Lindblad:Book}.

Consider the system initially prepared in a thermal state $\rho^{\text{ini}}_{\text{th}}$ at temperature $\beta$ and internal Hamiltonian $H_{\text{ini}}$. The system can then be driven from the thermal state $\rho^{\text{ini}}_{\text{th}}$ to another thermal state $\rho^{\text{fin}}_{\text{th}}$ with final Hamiltonian $H_{\text{fin}}$. This process can be done through a sequence of equilibrium states or through a far-from equilibrium process. In a non-equilibrium evolution, we decouple the system from the thermal bath and implement a unitary evolution with a time-dependent driving Hamiltonian $H(t)$, where the boundary conditions are $H(0) = H_{\text{ini}}$ and $H(\tau) = H_{\text{fin}}$. The final state achieved at time $\tau$ is  
denoted by $\rho(\tau)$. Then, from the non-equilibrium state $\rho(\tau)$, we drive the system to the thermal state associated with a bath at inverse temperature given by $\beta_{\text{fin}}$. 
During this process, an amount of irreversible entropy arises, reading
\begin{align}
\ve{\Sigma}  = \beta \left( \ve{w} - \Delta F \right) \text{ , } \label{Sigma1}
\end{align}
where $\Delta F$ is the Helmholtz free energy variation and $\ve{w}$ is the work realized on/by the system during the unitary evolution. In Ref.~\cite{Deffner:10}, it is shown 
that Eq.~(\ref{Sigma1}) can be expressed in terms of the relative entropy between the states $\rho(\tau)$ and $\rho^{\text{fin}}_{\text{th}}$. Namely, we have
\begin{align}
\ve{\Sigma} = \tr{\rho(\tau) \ln \rho(\tau)} - \tr{\rho(\tau)\ln \rho^{\text{fin}}_{\text{th}}} = \Dcal[\rho(\tau)||\rho^{\text{fin}}_{\text{th}}] \text{ . } \label{Sigma2}
\end{align}
On the other hand, by computing the total internal energy variation of the system we can write $\Delta U = \ve{w} + \ve{Q^{\text{th}}}$, where $\ve{Q^{\text{th}}}$ is the amount of 
heat exchanged between system and reservoir during the non-unitary process towards the thermal state at $\beta_{\text{fin}}$. Now, we can use the relation $\Delta U = \Delta F + (1/\beta)\Delta S$, where $\Delta S$ is the von Neumann entropy variation during the thermalization, to find
\begin{align}
\beta \left( \ve{w} - \Delta F \right) = \Delta S - \beta\ve{Q^{\text{th}}} \text{ , } \label{E3}
\end{align}
which is the standard expression for the (irreversible) entropy production adopted in Eq.~(\ref{ApTradS}) of the main text. 
Now, by using Eqs.~\eqref{Sigma1} and ~\eqref{Sigma2} in Eq.~(\ref{E3}), we obtain 
\begin{align}
\Delta S - \beta\ve{Q^{\text{th}}} = \Dcal[\rho(\tau)||\rho^{\text{fin}}_{\text{th}}] \text{ . }
\end{align}


\begin{thebibliography}{10}
	
	\bibitem{Geva:92}
	E.~Geva and R.~Kosloff,
	\newblock J. Chem. Phys. {\bf 96}, 3054 (1992).
	
	\bibitem{Lutz:16}
	O.~Abah and E.~Lutz,
	\newblock EPL (Europhysics Letters) {\bf 113}, 60002 (2016).
	
	\bibitem{Noah:10}
	N.~Linden, S.~Popescu, and P.~Skrzypczyk,
	\newblock Phys. Rev. Lett. {\bf 105}, 130401 (2010).
	
	\bibitem{Alicki:13}
	R.~Alicki and M.~Fannes,
	\newblock Phys. Rev. E {\bf 87}, 042123 (2013).
	
	\bibitem{Binder:15}
	F.~C. Binder, S.~Vinjanampathy, K.~Modi, and J.~Goold,
	\newblock New J. Phys. {\bf 17}, 075015 (2015).
	
	\bibitem{Kieu:04}
	T.~D. Kieu,
	\newblock Phys. Rev. Lett. {\bf 93}, 140403 (2004).
	
	\bibitem{Scully:03}
	M.~O. Scully, M.~S. Zubairy, G.~S. Agarwal, and H.~Walther,
	\newblock Science {\bf 299}, 862 (2003).
	
	\bibitem{Kosloff:84}
	R.~Kosloff,
	\newblock The Journal of chemical physics {\bf 80}, 1625 (1984).
	
	\bibitem{Rezek:06}
	Y.~Rezek and R.~Kosloff,
	\newblock New J. Phys. {\bf 8}, 83 (2006).
	
	\bibitem{Henrich:07}
	M.~J. Henrich, G.~Mahler, and M.~Michel,
	\newblock Phys. Rev. E {\bf 75}, 051118 (2007).
	
	\bibitem{Geusic:59}
	J.~Geusic, E.~Schulz-Du~Bois, R.~De~Grasse, and H.~Scovil,
	\newblock Journal of Applied Physics {\bf 30}, 1113 (1959).
	
	\bibitem{Quan:05}
	H.~T. Quan, P.~Zhang, and C.~P. Sun,
	\newblock Phys. Rev. E {\bf 72}, 056110 (2005).
	
	\bibitem{Lutz:16-Science}
	J.~Ro{\ss}nagel {\em et~al.},
	\newblock Science {\bf 352}, 325 (2016).
	
	\bibitem{Scovil:59}
	H.~Scovil and E.~Schulz-DuBois,
	\newblock Phys. Rev. Lett. {\bf 2}, 262 (1959).
	
	\bibitem{Peterson:18}
	J.~P. Peterson {\em et~al.},
	\newblock arXiv preprint arXiv:1803.06021  (2018).
	
	\bibitem{Klatzow:17}
	J.~Klatzow {\em et~al.},
	\newblock Phys. Rev. Lett. {\bf 122}, 110601 (2019).
	
	\bibitem{Nielsen:Book}
	M.~A. Nielsen and I.~L. Chuang,
	\newblock {\em Quantum Computation and Quantum Information: 10th Anniversary
		Edition}, 10th ed. (Cambridge University Press, New York, NY, USA, 2011).
	
	\bibitem{PRA.78.Davidovich}
	A.~Salles {\em et~al.},
	\newblock Phys. Rev. A {\bf 78}, 022322 (2008).
	
	\bibitem{PRL.99.Topo}
	C.~E.~R. Souza, J.~A.~O. Huguenin, P.~Milman, and A.~Z. Khoury,
	\newblock Phys. Rev. Lett. {\bf 99}, 160401 (2007).
	
	\bibitem{PRA.82.Borges}
	C.~V.~S. Borges, M.~Hor-Meyll, J.~A.~O. Huguenin, and A.~Z. Khoury,
	\newblock Phys. Rev. A {\bf 82}, 033833 (2010).
	
	\bibitem{NaturePhot.Kagalwala-2012}
	K.~H. Kagalwala, G.~Di~Giuseppe, A.~F. Abouraddy, and B.~E.~A. Saleh,
	\newblock Nature Photonics {\bf 7} (2012).
	
	\bibitem{Opt.Lett.Balthazar-2016}
	W.~F. Balthazar {\em et~al.},
	\newblock Optics Letters {\bf 41}, 5797 (2016).
	
	\bibitem{PRA.77.Cadu-Cripto}
	C.~E.~R. Souza {\em et~al.},
	\newblock Phys. Rev. A {\bf 77}, 032345 (2008).
	
	\bibitem{PRA.83.Zela-Teleport}
	A.~Z. Khoury and P.~Milman,
	\newblock Phys. Rev. A {\bf 83}, 060301 (2011).
	
	\bibitem{Opt.Exp.Souza-Cnot-2010}
	C.~E.~R. Souza and A.~Z. Khoury,
	\newblock Optics Express {\bf 18}, 9207 (2010).
	
	\bibitem{JOPS-B.Cod.Op.Balthazar-2016}
	W.~F. Balthazar and J.~A.~O. Huguenin,
	\newblock Journal of the Optical Society of America B {\bf 33}, 1649 (2016).
	
	\bibitem{PRA.97.Enviro-Passos}
	M.~H.~M. Passos {\em et~al.},
	\newblock Phys. Rev. A {\bf 97}, 022321 (2018).
	
	\bibitem{Talkner:07}
	P.~Talkner, E.~Lutz, and P.~H\"anggi,
	\newblock Phys. Rev. E {\bf 75}, 050102 (2007).
	
	\bibitem{Alicki:79}
	R.~Alicki,
	\newblock J. Phys. A: Math. Gen. {\bf 12}, L103 (1979).
	
	\bibitem{Anders:13}
	J.~Anders and V.~Giovannetti,
	\newblock New J. Phys. {\bf 15}, 033022 (2013).
	
	\bibitem{Lutz:18}
	O.~Abah and E.~Lutz,
	\newblock Phys. Rev. E {\bf 98}, 032121 (2018).
	
	\bibitem{Photonic-State-Tomography}
	J.~B. Altepeter, E.~R. Jeffrey, and P.~G. Kwiat,
	\newblock Adv. Atom. Mol. Opt. Phy. {\bf 52}, 105 (2005).
	
	\bibitem{Jones-i:41}
	R.~C. Jones,
	\newblock J. Opt. Soc. Am. {\bf 31}, 488 (1941).
	
	\bibitem{Procaccia:76}
	I.~Procaccia and R.~Levine,
	\newblock J. Chem. Phys. {\bf 65}, 3357 (1976).
	
	\bibitem{Jarzynski:97}
	C.~Jarzynski,
	\newblock Phys. Rev. Lett. {\bf 78}, 2690 (1997).
	
	\bibitem{Parrondo:09}
	J.~M. Parrondo, C.~Van~den Broeck, and R.~Kawai,
	\newblock New J. Phys. {\bf 11}, 073008 (2009).
	
	\bibitem{Lindblad:Book}
	C.~Lindblad,
	\newblock {\em Non-equilibrium entropy and irreversibility} (Reidel Publishing
	Company, Dordrecht, Holland, 1983).
	
	\bibitem{Batalhao:15}
	T.~B. Batalh\~ao {\em et~al.},
	\newblock Phys. Rev. Lett. {\bf 115}, 190601 (2015).
	
	\bibitem{Adolfo:14}
	A.~del Campo, J.~Goold, and M.~Paternostro,
	\newblock Sci. Rep. {\bf 4}, 6208 (2014).
	
	\bibitem{Plastina:14}
	F.~Plastina {\em et~al.},
	\newblock Phys. Rev. Lett. {\bf 113}, 260601 (2014).
	
	\bibitem{Brask:15}
	J.~B. Brask and N.~Brunner,
	\newblock Phys. Rev. E {\bf 92}, 062101 (2015).
	
	\bibitem{Haupt:14}
	F.~Haupt, A.~Imamoglu, and M.~Kroner,
	\newblock Phys. Rev. Applied {\bf 2}, 024001 (2014).
	
	\bibitem{Hofer:17}
	P.~P. Hofer, J.~B. Brask, M.~Perarnau-Llobet, and N.~Brunner,
	\newblock Phys. Rev. Lett. {\bf 119}, 090603 (2017).
	
	\bibitem{Tung:Book}
	W.-K. Tung,
	\newblock {\em Group Theory in Physics} (World Scientific, 1985).
	
	\bibitem{Deffner:10}
	S.~Deffner and E.~Lutz,
	\newblock Phys. Rev. Lett. {\bf 105}, 170402 (2010).
	
	\bibitem{Apollaro:15}
	T.~J. Apollaro, G.~Francica, M.~Paternostro, and M.~Campisi,
	\newblock Phys. Scr. {\bf 2015}, 014023 (2015).
	
\end{thebibliography}

\end{document}